%% file: main.tex
\documentclass[conference]{IEEEtran}

\usepackage{epsfig}
\usepackage{graphicx}
\usepackage{color}
\usepackage{amsmath}%
\usepackage{mathrsfs}
\usepackage{pgfplots}
\usepackage{tikz}
\usetikzlibrary{shapes,shadows,arrows,calc,plotmarks,intersections}
\usepackage{caption}
\usepackage{makecell}

\usepackage{amsthm} 
\usepackage{amsmath} 
\usepackage{amssymb} 
\usepackage{graphics} 
\usepackage{graphicx} 
\usepackage{setspace} 
\usepackage{datetime} 
\usepackage{subfigure}
\usepackage{algorithm}
\usepackage{algpseudocode}
\usepackage{caption}
\usepackage{float}
\usepackage{listings}
\usepackage{makecell}
\usepackage{framed}
\usepackage{wrapfig}
\usepackage{rotating}
\usepackage{graphicx}
\usepackage{caption}
\usepackage{amsmath}
\usepackage{booktabs}
\usepackage{amssymb}
\usepackage{graphicx}
\usepackage{mathrsfs}
\usepackage{mathtools}
\usepackage{epsfig}
\usepackage{fancyhdr}
\usepackage{epsfig}
\usepackage{graphicx}
\usepackage{color}
\usepackage{amsmath}%
\usepackage{mathrsfs}
\usepackage{pgfplots}
\usepackage{multirow}
\usepackage[square, numbers, comma, sort&compress]{natbib}
\ifCLASSINFOpdf

\else
 
\fi

\begin{document}
%
\title{Linear Prediction based Data Detection of Convolutional Coded DQPSK in SIMO-OFDM }

\author{\IEEEauthorblockN{Vineel K Veludandi}
\IEEEauthorblockA{Dept. of Electrical Engineering\\
Indian Institute of Technology Kanpur\\
Kanpur, India 208016\\
Email: vineelkv@iitk.ac.in}
\and

\IEEEauthorblockN{K Vasudevan}
\IEEEauthorblockA{Dept. of Electrical Engineering\\
Indian Institute of Technology Kanpur\\
Kanpur, India 208016\\
Email: vasu@iitk.ac.in}}

\maketitle

\begin{abstract}
Data detection of convolutional coded differential quaternary phase shift keyed (DQPSK) signals using a predictive Viterbi algorithm (VA) based receiver, is presented for single input, multiple output - orthogonal frequency division multiplexed (OFDM) systems. The receiver has both error correcting capability and also the ability to perform channel estimation (prediction). The predictive VA operates on a supertrellis with just $S_{\mathrm{ST}}=S_{\mathrm{E}}\times 2^{P-1}$ states instead of $S_{\mathrm{ST}}=S_{\mathrm{E}}\times 2^{P}$ states, where the complexity reduction is achieved by using the concept of isometry (here $S_{\mathrm{E}}$ denotes the number of states in the encoder trellis and $P$  denotes the prediction order). Though the linear prediction based data detection in turbo coded OFDM \cite{vin_wpc} and the bit interleaved coded (BIC) OFDM \cite{app_based2} systems perform better than the proposed approach in terms of bit error rate (BER) for a given signal to noise ratio (SNR), the decoding delay of the proposed approach is significantly lower than that of the BIC and the turbo coded OFDM systems.  
\end{abstract}


\IEEEpeerreviewmaketitle

\section{Introduction}
OFDM has the ability to convert a frequency selective fading channel into a frequency flat channel \cite{text_book}. Though the coherent detectors perform better than the linear prediction (LP)-based detectors in terms of BER \cite{vin_wpc,elsevier,vineel_va}, for a given SNR, coherent detectors are not throughput efficient since pilots have to be transmitted in every OFDM frame for the purpose of estimating the channel frequency response \cite{pilot_based1,pilot_based2,pilot_based3}. The throughput is defined as the ratio of the number of data symbols to the total symbols in an OFDM frame. However, since LP-based detectors perform data detection by using the channel statistics alone, LP-based detectors are throughput efficient since the statistics of the channel need to be estimated only once during the first OFDM frame \cite{vin_wpc}. In wireless communications, it is assumed that the channel statistics remain same for every OFDM frame but the channel impulse and frequency response changes with every OFDM.

  In the proposed approach of LP-based data detection in convolutional coded SIMO-OFDM signals, the predictive VA operates on a supertrellis \cite{super_trellis1,super_trellis2,vin_wpc,sivp,text_book_soft}, obtained by combining the memory of the convolutional encoder and a prediction filter. A rate-1/2 convolutional encoder with $S_{\textrm{E}}$ states when combined with a prediction filter of order $P$ would result in a supertrellis with $S_{\mathrm{E}}\times 2^{P}$ states. However, the complexity is reduced to just $S_{\mathrm{E}}\times 2^{P-1}$ states using the concept of isometry \cite{elsevier,vin_wpc}. The channel prediction filter exploits the high degree of correlation in the channel frequency response at the output of the fast Fourier transform (FFT) in the OFDM receiver when the length of the channel impulse response is much smaller than the FFT length. Perfect timing and carrier synchronization is assumed. Simulation results are compared against the ideal coherent detector where perfect channel-state information (CSI) is assumed. It is shown that the LP-based receiver performs close to the ideal coherent receiver. Though the BER performance of the LP-based detection in bit interleaved coded (BIC) OFDM \cite{app_based2} and the turbo coded OFDM systems \cite{vin_wpc} perform better than the proposed approach, the decoding delay of the proposed approach is significantly lower than that of the BIC and the turbo coded OFDM systems.
  
  This paper is organized as follows. The notation used throughput this paper is given in Section 2. The system model is given in Section 3. The proposed linear prediction-based receiver is discussed in Section 4. In Section 5, we give the simulation results. Finally in Section 6, we give the conclusion and the scope for the future work.
  
\section{Notation}
In this paper, all lower-case and upper-case letters without a tilde e.g. $g_k$ represent real-valued scalar. Letters with a tilde e.g. $\tilde h_k$, denote complex quantities. However, complex symbols are denoted by $S_k$ (without a tilde).  Boldface letters represent vectors or matrices. All letters with a hat, e.g. $\hat{X}_k$  denote the statistical estimate of $\tilde{X}_k$  (or $X_k$, if it is real-valued). The $(\cdot)^*$  denotes complex conjugate, $(\cdot)^{\mathscr{H}}$ denotes conjugate transpose and $E\left[\cdot\right]$ denotes the expectation operation. We also assume that bit $0$ maps to $+1$ and bit $1$ maps to $-1$.  

\section{System Model}
\subsection{Transmitter}
\begin{figure*}[h]
\centering
\input{predictive_va.pstex_t}
\caption{Block diagram for the convolutioinal coded OFDM system. Two consecutive bits of $b_k$ are mapped to $S_k$ using the differential encoding rules given in Table \ref{tab:dif_lit1}.}
\label{block_diagram}
\end{figure*}
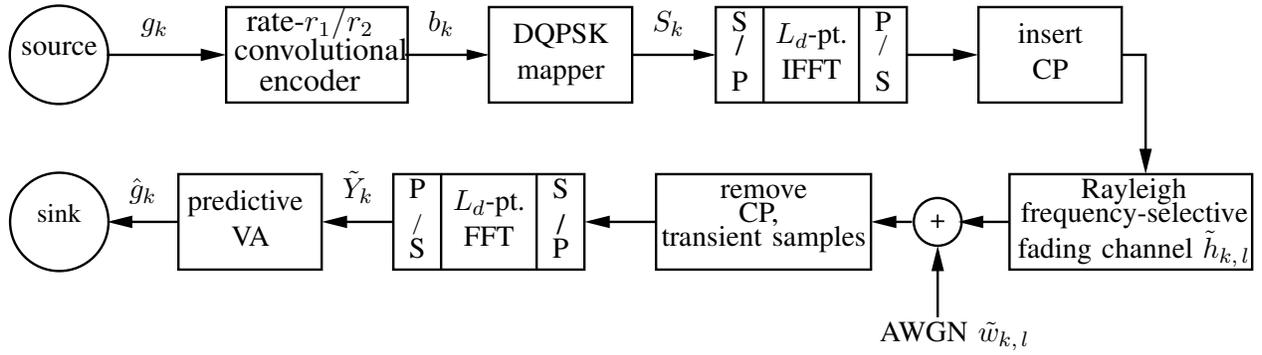

 The binary input data $g_k$ ($0 \le k \le L_d/2 - 1$) from the source are encoded using a rate-$r_1/r_2$ convolutional encoder. The encoded data $b_k$ is mapped to DQPSK according to the differential encoding rules \cite{text_book} given in Table \ref{tab:dif_lit1} to get $S_k$. The symbol stream $S_k$ is fed to a serial to parallel converter (S/P) and loaded on to the OFDM sub-carriers by an $L_d$-point inverse fast Fourier transform (IFFT) operation. The length of the cyclic prefix (CP) is equal to the length of the channel memory $L_{\textrm{CP}} = L_h-1$ \cite{text_book}, and is inserted into the OFDM frame.  Note that the overall rate of the transmitter in Figure \ref{block_diagram} is $1$, that is, one bit is sent per transmission. For every data bit, one coded QPSK symbol is transmitted, hence one bit of information is sent per transmission. 
 
 \begin{table}[H]
\fontsize{10}{10}\selectfont 
\centering
\caption{Differential encoding rules \cite{text_book}}
\label{tab:dif_lit1}       
\begin{tabular}{ccc}
\toprule
\makecell{Dibit \\ $\left(b_{k-1}b_{k}\right)$} & \makecell{Decimal equivalent of\\the dibit ($\mathscr{S}_{0,\,j}$)} & \makecell{Phase change \\ (in radians)} \\
\midrule
$00$ & 0 &  $0$  \\
$01$ & 1 & $\pi/2$  \\
$10$ & 2 &  $3\pi/2$  \\
$11$ & 3 &  $\pi$  \\
\bottomrule
\end{tabular}
\end{table}
\subsection{Channel Model}

 We assume a Rayleigh frequency selective fading channel having a uniform power delay profile \cite{indouk,vineel_va,shimla,wpc,spain,vin_wpc}. Though an exponential power delay profile is more practical, we expect the uniform power delay profile to give the worst case BER performance since all channel taps (intersymbol interference (ISI) terms) have the same power. The channel is assumed to be time-invariant over each OFDM frame  and varies independently from frame to frame i.e. quasistatic. For the $l^{\mathrm{th}}$ diversity arm ($0\le l \le N_r-1$), the channel impulse response $\tilde h_{k,l}$ ($0 \le k \le L_h-1$) and AWGN noise $\tilde w_{k,l}$ ($0 \le k \le L_f-1$)  are both wide-sense stationary (WSS) circularly symmetric complex Gaussian random variables with autocorrelation given by:
\begin{eqnarray}
\frac{1}{2}{E}{\left[{\tilde h_{k,l}}{\tilde h^{*}_{k',l'}} \right]} &= \begin{cases}
\sigma^{2}_{f},& \text{if} $ {\it k=k'} \ and {\it\ l=l'} $\; \nonumber \\ 
0,& \text{otherwise} \; 
\end{cases}\\
\frac{1}{2}{E}{\left[{\tilde w_{k,l}}{\tilde w^{*}_{k',l'}} \right]} &= \begin{cases}
\sigma^{2}_{w},& \text{if} $ {\it k=k'} \ and {\it\ l=l'} $\; \\
0,& \text{otherwise}. \;
\end{cases}
\end{eqnarray}

\section{Receiver}
\label{receiver}
The output of the FFT operation at the $l^{\mathrm{th}}$ diversity arm in the receiver is given by:
\begin{align}
\label{equation:system_model}
\tilde Y_{k,\,l}= \tilde H_{k,\,l} S_{k}  + \tilde W_{k,\,l},\quad 0 \le k \le L_f - 1, \ 0\le l \le N_r-1 
\end{align}
where
\begin{align}
 \tilde H_{k,\,l} &= \sum\limits_{i=0}^{L_{h}-1} {\tilde{h}_{i,\,l}} e^{{-\mathrm{j}2\pi ik/L_{d}}} \nonumber \\ 
\tilde W_{k,\,l} &= \sum\limits_{i=0}^{L_{f}-1} {\tilde{w}_{i,\,l}} e^{{-\mathrm{j}2\pi ik/L_{d}}}.
\end{align}
The 1-D autocorrelation of the discrete Fourier transform (DFT) of the AWGN samples is \cite{shimla}
\begin{align}
\label{equ:awgn_autocorr}
 \frac{1}{2}E\left[{\tilde W_{k,\,l} \tilde W^*_{k-m,\,l}}\right] = L_d\sigma_w^2\delta _{K,\,m}.
\end{align}  
where $\delta_{K,\,m}$ is the Kronecker delta function defined as:
\begin{align}
\label{equation:Kronecker}
\delta _{K,\,m} = \begin{cases}
1,& \text{if} \ m = 0\; ,\\
0,& \text{if} \ m \neq 0. \;
\end{cases}
\end{align}

The autocorrelation of $\tilde H_{k,l}$ is given by \cite{shimla}:
\begin{align}
\label{equ:define_autocorr}
\tilde{R}_{\tilde H \tilde H,\,m} \buildrel \Delta \over =  \frac{1}{2} E\left[ {\tilde{H}_{k,\,l}\tilde{H}^{*}_{k-m,\,l}} \right] = \sigma ^{2}_{f}\sum \limits _{n=0}^{L_{h}-1}e^{-\mathrm{j}2\pi nm/L_{d}}.
\end{align}

Now consider
\begin{align}
\label{equ:define_X}
\tilde{X}_{k,\,l}=\tilde{Y}_{k,\,l}/S_k.
\end{align}
During data detection, the symbols $S_k$ are obtained from the supertrellis. 
 
 The autocorrelation of $\tilde{X}_{k,\,l}$ is given by:
\begin{align}
\label{equ:autocor}
\tilde{R}_{\tilde{X}\tilde{X},\,m} \buildrel \Delta \over = \frac{1}{2}E\left[\tilde{X}_{k,\,l} \tilde{X}^*_{k-m,\,l} \right] = \tilde{R}_{\tilde{H}\tilde{H},\,m} + \frac{\sigma^2_w L_d}{\left| S_k\right|^2}\delta_{K,\,m} 
 \end{align}
where we have assumed that
\begin{align}
\left| S_k\right|^2 = \mathrm{constant}.
\end{align}  
The key idea behind this approach is to estimate (predict) $\tilde X_{k,\,l}$ assuming $S_{k}$ is known, as follows \cite{text_book}:
\begin{align}
\label{equ:key_idea}
\hat{X}_{k,\,l} = -\sum \limits ^P_{j=1}\tilde{a}_{P,\,j}  \tilde{X}_{k-j,\,l}
\end{align}
where $\tilde{a}_{P,\,j}$ denotes the  $j^{\mathrm{th}}$ coefficient of the optimum $P^{th}$-order predictor. Note that $\tilde X_{k,\,l}\approx \tilde H_{k,\,l}$ at high SNR.

The prediction error is \cite{text_book}:
\begin{align}
\tilde{z}_{k,\,l} \buildrel \Delta \over =  \tilde{X}_{k,\,l} -\hat{X}_{k,\,l} = \sum \limits ^P_{j=0}\tilde{a}_{P,\,j} \tilde{X}_{k-j,\,l}
\end{align}
and the 1-D prediction error variance is given by \cite{text_book}:
\begin{align}
\label{equ:error_variance}
\sigma^2_{e,\,P}   \buildrel \Delta \over = \frac{1}{2}{E}\left[\left|\tilde{z}_{k,\,l} \right|^2 \right]  =  \sum \limits ^P_{j=0}\tilde{a}_{P,\,j}\tilde{R}_{\tilde{X}\tilde{X},\,-j}
\end{align}
where $\tilde{a}_{P,\,0}=1$.  In Section \ref{pred_principle} we give a formal derivation of the linear prediction-based receiver.

In practice of course, the autocorrelation $\tilde{R}_{\tilde{X}\tilde{X},\,m}$ required for generating the prediction filter coefficients is not known. Hence the autocorrelation needs to be estimated only once (see Section 4.1 in \cite{vin_wpc}). However, in this paper we assume that the receiver has perfect knowledge of the channel and noise statistics.
\subsection{The Suboptimal Predictive Maximum Likelihood (ML) decoder \cite{elsevier}}
\label{pred_principle}
The received signal at the $l^{\mathrm{th}}$ diversity arm can be represented as: 
\begin{align}
\mathbf{\tilde Y}_l&= \mathbf{ S}^{\left(q\right)} \mathbf{\tilde H}_l  + \mathbf{ \tilde W}_l,\; 0\leq q \leq M^{L_d}-1,\; 0\leq l\leq N_r-1
\end{align}
where $\mathbf{\tilde Y}_l$ is an $L_d \times 1$ column vector of received samples, $\mathbf{ S}^{\left(q\right)}$ is an $L_d \times L_d$ diagonal matrix with elements containing the $q^{\mathrm{th}}$ possible QPSK symbol sequence, $\mathbf{\tilde H}_l$ is an $L_d \times 1$ column vector of the channel DFT and $\mathbf{ \tilde W}_l$ is an $L_d \times 1$ column vector containing the DFT of the AWGN samples $\tilde{w}_{k,\,l}$ in Figure \ref{block_diagram}, and $M=4$ (QPSK constellation).

The ML detector decides in favour of $\mathbf{ S}^{\left(q\right)}$ that maximizes the joint conditional pdf 
\begin{align}
\label{equation:map_rule1}
 \mathop {\mathrm{max}}\limits_{q}\; & p\left(\mathbf{\tilde Y}_0, \mathbf{\tilde Y}_1, \hdots, \mathbf{\tilde Y}_{N_{r}-1}|\mathbf{ S}^{\left(q\right)}\right) \nonumber \\
\Rightarrow \; \mathop {\mathrm{max}}\limits_{q}\; & \prod \limits_{l=0}^{N_r-1} p\left( \mathbf{\tilde Y}_l | \mathbf{ S}^{\left(q\right)} \right)
\end{align}
where $p\left(\cdot\right)$ denotes the probability density function, and we have assumed that $\mathbf{\tilde H}_l$, $\mathbf{ \tilde W}_l$ and hence $\mathbf{\tilde Y}_l$ are independent over $l$. Ignoring constants and substituting for the conditional pdfs in (\ref{equation:map_rule1}), we get
\begin{align}
\label{equ:map_rule2}
 \mathop {\mathrm{max}}\limits_{q}\; \exp\left({-\frac{1}{2}\sum \limits _{l=0}^{N_r-1}\mathbf{\tilde Y}_l^{\mathscr{H}}\left(\mathbf{\tilde R}^{\left(q\right)}\right)^{-1}\mathbf{\tilde Y}_l}\right)
\end{align}
where
\begin{align}
\mathbf{\tilde R}^{\left(q\right)} &\buildrel \Delta \over = \frac{1}{2}{E}\left[ \mathbf{\tilde Y}_l \mathbf{\tilde Y}_l^{\mathscr{H}}|\mathbf{ S}^{\left(q\right)} \right] \nonumber \\  &= \frac{1}{2} \mathbf{ S}^{\left(q\right)} {E}\left[ \mathbf{\tilde H}_l \mathbf{\tilde H}_l^{\mathscr{H}}\right] \left(\mathbf{ S}^{\left(q\right)}\right)^{\mathscr{H}} + \sigma_w^2L_d\mathbf{I} \nonumber \\
&\approx \frac{1}{2} \mathbf{ S}^{\left(q\right)} {E}\left[ \mathbf{\tilde H}_l \mathbf{\tilde H}_l^{\mathscr{H}}\right] \left(\mathbf{ S}^{\left(q\right)}\right)^{\mathscr{H}} \; (\textrm{at high SNR})\nonumber \\ 
&=  \mathbf{ S}^{\left(q\right)} \mathbf{\Phi} \left(\mathbf{ S}^{\left(q\right)}\right)^{\mathscr{H}}  \; (\textrm{say})
\end{align}
where we have used (\ref{equ:awgn_autocorr}) and (\ref{equ:define_autocorr}). Now, by applying Cholesky decomposition of the autocovariance matrix $\mathbf{\Phi}$, it can be shown that \cite{text_book}
\begin{align}
{\mathbf{\Phi}}^{-1} = \mathbf{\tilde{B}}^{\mathscr{H}}\mathbf{D}^{-1}\mathbf{\tilde{B}}
\end{align}
where 
\begin{align}
\label{equation:B_matrix}
\mathbf{\tilde{B}}\buildrel \Delta \over =\left[ {\begin{array}{*{20}{c}}
{1}&{0}&{\hdots}&{0}\\
{\tilde{a}_{1,\,1}}&{1}&{\hdots}&{0}\\
{\vdots}&{\vdots}&{\vdots}&{\vdots}\\
{\tilde{a}_{L_d-1,\,L_d-1}}&{\tilde{a}_{L_d-1,\,L_{d}-2}}&{\hdots}&{1}
\end{array}} \right]
\end{align}
is the ($L_d \times L_d$) matrix of predictor coefficients with $\tilde a_{i,\,\tau}$ being the $\tau^{\mathrm{th}}$ coefficient of the optimum $i^{\mathrm{th}}$-order predictor and the ($L_d \times L_d$) matrix
\begin{align}
\label{equation:D_matrix}
\mathbf{D}\buildrel \Delta \over = \left[ {\begin{array}{*{20}{c}}
{\sigma_{e,\,0}^2}&{\hdots}&{0}\\
{\vdots}&{\ddots}&{\vdots}\\
{0}&{\hdots}&{\sigma_{e,\,L_d-1}^2}
\end{array}} \right]
\end{align}
where $\sigma _{e,\,k}^2$ is the 1-D prediction error variance of the optimum $k^{\mathrm{th}}$-order predictor as given by (\ref{equ:error_variance}) and $\sigma _{e,\,0}^2= \tilde{R}_{\tilde{H}\tilde{H},\,0}  \approx  \tilde{R}_{\tilde{X}\tilde{X},\,0} $ at high SNR.

Now, the maximization rule in (\ref{equ:map_rule2}) can be expressed as
\begin{align}
\label{equation:map_rule}
\mathop {\mathrm{min}}\limits_{q}\;&  \sum \limits _{l=0}^{N_r-1} \mathbf{\tilde Y}_l^{\mathscr{H}} \left( \left(\mathbf{ S}^{\left(q\right)}\right)^{\mathscr{H}} \right)^{-1} \mathbf{B}^{\mathscr{H}}\mathbf{D}^{-1}\mathbf{B} \left(\mathbf{ S}^{\left(q\right)}\right)^{-1} \mathbf{\tilde Y}_l \nonumber \\
  \Rightarrow \; &\mathop {\mathrm{min}}\limits_{q}\; \sum \limits _{l=0}^{N_r-1}  \sum \limits_{k=0}^{L_d-1} \frac{\left| \tilde{z}_{k,\,l}^{{\left(q\right)}} \right|^2 }{2\sigma^2_{e,\,k}} 
\end{align}
where the prediction error $\tilde{z}_{k,\,l}^{\left(q\right)}$ is an element of 
\begin{align}
\left[ {\begin{array}{*{20}{c}}
{\tilde{z}_{0,\,l}^{\left(q\right)}}\\
{\tilde{z}_{1,\,l}^{\left(q\right)}}\\
{\vdots}\\
{\tilde{z}_{L_d-1,\,l}^{\left(q\right)}}
\end{array}} \right] \buildrel \Delta \over = \mathbf{\tilde{z}}^{\left(q\right)}=\mathbf{B} \left(\mathbf{ S}^{\left(q\right)}\right)^{-1} \mathbf{\tilde Y}_l.
\end{align}
Assuming that a $P^{\mathrm{th}}$-order predictor completely decorrelates noise, (\ref{equation:map_rule}) can be written as 
\begin{align}
\label{equation:final_map_rule}
 \mathop {\mathrm{min}}\limits_{q}\; \sum \limits _{l=0}^{N_r-1}  \sum \limits_{k=0}^{P-1} \frac{\left| \tilde{z}_{k,\,l}^{\left(q\right)} \right|^2 }{2\sigma^2_{e,\,k}} +\sum \limits _{l=0}^{N_r-1}  \sum \limits_{k=P}^{L_d-1} \frac{\left| \tilde{z}_{k,\,l}^{\left(q\right)} \right|^2 }{2\sigma^2_{e,\,P}} . 
\end{align}
Note that the first double summation in (\ref{equation:final_map_rule}) denotes the ``transient" part and the second double summation denotes the ``steady state" part. Observe also that the predictor coefficients in (\ref{equation:B_matrix}) correspond to the autocorrelation of the channel frequency response. In practice, the predictor coefficients are obtained from the autocorrelation of $\tilde{X}_{k,\,l}$. Finally we note that the complexity in (\ref{equation:final_map_rule}) increases exponentially with $L_d$.  In the Subsection \ref{sec:pred_VA_ALG}, we present the predictive VA, whose complexity increases linearly with $L_d$.

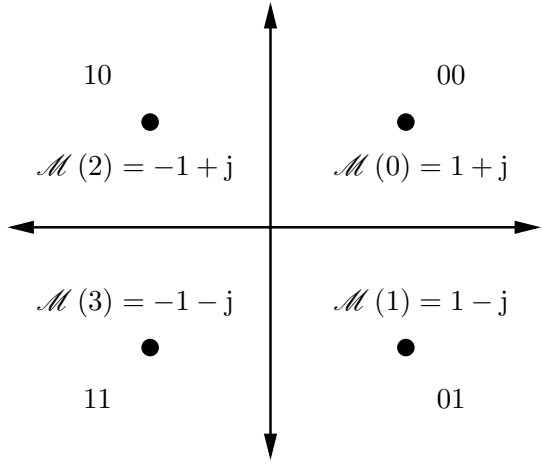
\begin{figure}
\centering
\input{gray_arxiv.pstex_t}
\caption{Encoding rules for the QPSK constellation \cite{text_book}}
\label{mapping}
\end{figure}

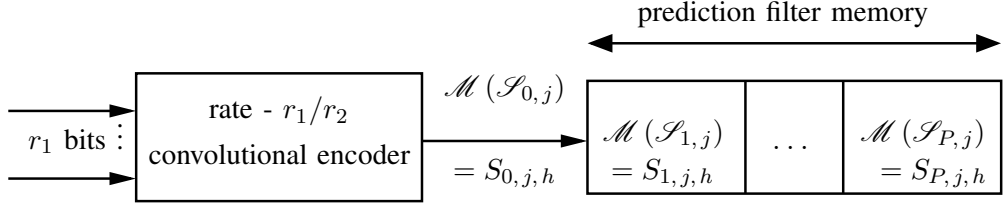
\begin{figure*}
\centering
\input{Fig4_arxiv.pstex_t}
\caption{Procedure for constructing the supertrellis \cite{text_book_soft}}
\label{fig:supertrellis}
\end{figure*}

\subsection{Supertrellis Construction \cite{text_book_soft,vin_wpc}}
\label{section:supertrellis}

In Section \ref{receiver} (just after (\ref{equ:define_X})), we mentioned that the symbols $S_k$ are not known, and in practice they are obtained from a supertrellis. Consider the predictive VA in Figure \ref{block_diagram}. The inner decoder trellis must be modified to a supertrellis which incorporates the memory of the prediction filter. 

Assume that the $r_2$ coded bits from the inner rate-${r_1}/{r_2}$ convolutional encoder (in this work $r_1=1$,\, $r_2=2$) are mapped\footnotemark \footnotetext{In this section we assume that the DQPSK mapper in Figure \ref{block_diagram} is absent. The need to have a DQPSK mapper will be explained in Section \ref{comp_red_lit}.} to an $M$-ary ($M=2^{r_2}$)  constellation according to the set partitioning rule, e.g. $ S_{0,\,j,\,h} = {\mathscr M}\left(\mathscr{S}_{0,\,j}\right)$ \cite{text_book} (see Figure \ref{mapping}). Here, $\mathscr{S}_{0,\,j}$ $\left(0\leq \mathscr{S}_{0,\,j} \leq 2^{r_2}-1\right)$ is the decimal equivalent of the $n$ coded bits $b_kb_{k-1}\hdots b_{k-n+1}$ in Figure \ref{block_diagram}, and is referred to as the input code digit. 
Note that since the supertrellis is a periodic structure, we have removed the subscript $k$ in $S_k$ and replaced it with $S_{i,\,j,\,h}$. The subscript ``$i$" in $S_{i,\,j,\,h}$ refers to the $i^{\mathrm{th}}$ memory element of the prediction filter (the $0^{\mathrm{th}}$ element is the input), the subscript ``$j$" refers to the present supertrellis state and ``$h$" denotes the next supertrellis state. Observe that the symbol sequence $S_k$ in Figure \ref{block_diagram} corresponds to one of the paths through the supertrellis.

Now consider Figure \ref{fig:supertrellis}. Note that
\begin{align}
\label{equation:sstate_contents}
\lbrace\mathscr{M}\left(\mathscr{S}_{0,\,j}\right), \mathscr{M}\left( \mathscr{S}_{1,\,j}\right),\hdots,\mathscr{M}\left(\mathscr{S}_{P,\,j}\right)\rbrace
\end{align}
in Figure \ref{fig:supertrellis} is a valid encoded symbol sequence. In (\ref{equation:sstate_contents}), the subscript $P$ refers to a $P^{\mathrm{th}}$-order prediction filter and $j$ refers to the $j^{\mathrm{th}}$ supertrellis state, as will be explained later.

Let $S_{\mathrm{E}}$ denote the number of encoder states. For any given present convolutional encoder state $\mathscr{E}_i$ $\left(0\leq  i \leq S_{\textrm{E}}-1 \right)$, there are $2^{r_1}=N$ possible encoded symbols. Hence, starting from any particular encoder state, there are $N^{P}$ ways in which a prediction filter of order $P$ can be populated (see Figure \ref{fig:supertrellis}). Therefore, the total number of ways in which a $P^{\mathrm{th}}$-order predictor can be populated is $S_{\mathrm{E}}\times N^{P}$ which is also equal to the number of supertrellis states. Therefore 
\begin{align}
S_{\mathrm{ST}}=S_{\mathrm{E}}\times N^{P}.
\end{align}
Let $\mathscr{F}_{m}$ $\left(0 \leq m \leq N^P-1 \right)$ denote the prediction filter state. Supertrellis state is given by $\mathscr{S}_{\textrm{ST},\,j}$, where
\begin{align}
j=i \times N^P + m, \quad 0\leq j \leq S_{\textrm{E}} \times N^P-1.
\end{align}
Symbolically, the supertrellis state can be represented as:
\begin{align}
\label{super_sym_not}
\mathscr{S}_{\textrm{ST},\,j}: \lbrace \mathscr{E}_i;\mathscr{F}_m\rbrace.
\end{align}
Let us represent the prediction filter state $\mathscr{F}_m$ by an $N$-ary $P$-tuple as follows:
\begin{align}
\label{equation:n_ary_tuple}
\mathscr{F}_m: \lbrace \mathscr{N}_{1,\,m} \hdots \mathscr{N}_{P,\,m} \rbrace
\end{align}
where the input digits are denoted by
\begin{align}
\mathscr{N}_{t,\,m} \in \lbrace0,\,\hdots,\,N-1  \rbrace  \quad \textrm{for} \; 1\leq t \leq P
\end{align}
such that
\begin{align}
m = \sum \limits_{t=1}^{P}\mathscr{N}_{P+1-t,\,m}{N}^{t-1}
\end{align}
is the decimal equivalent of the $N$-ary, $P$-tuple in (\ref{equation:n_ary_tuple}).

 $\mathscr{F}_m$ is actually the input sequence to the encoder in Figure \ref{fig:supertrellis} with $\mathscr{N}_{P,\,m}$ being the initial input digit. Let $\mathscr{E}_s$ $\left(0\leq s\leq S_{\textrm{E}}-1 \right)$ be the encoder state corresponding to the input digit $\mathscr{N}_{P,\,m}$. The code digit sequence corresponding to the supertrellis state $\mathscr{S}_{\textrm{ST},\,j}$ is generated as follows:
\begin{align}
&\mathscr{E}_s, \mathscr{N}_{P,\,m}  \rightarrow \mathscr{E}_a, \mathscr{S}_{P,\,j} \quad \textrm{for}\; 0\leq a < S_{\textrm{E}},\;0\leq  \mathscr{S}_{P,\,j} < M \nonumber \\
&\mathscr{E}_a, \mathscr{N}_{P-1,\,m} \rightarrow \mathscr{E}_b, \mathscr{S}_{P-1,\,j} \; \textrm{for}\; 0\leq b < S_{\textrm{E}},\;0\leq  \mathscr{S}_{P-1,\,j} < M 
\end{align}
which means: the encoder at (starting) state $\mathscr{E}_s$ with input digit $\mathscr{N}_{P,\,m}$ yields the code digit $\mathscr{S}_{P,\,j}$ and the next encoder state $\mathscr{E}_a$ and so on. We repeat this procedure till the last input digit, to get: 
\begin{align}
\label{equation:super_final}
\mathscr{E}_c, \mathscr{N}_{1,m} \rightarrow \mathscr{E}_i, \mathscr{S}_{1,\,j} \quad \textrm{for}\; 0\leq c < S_{\textrm{E}},\;0\leq  \mathscr{S}_{1,\,j} < M \nonumber \\
\mathscr{E}_i,  \mathscr{N}_{0,\,m} \rightarrow \mathscr{E}_f, \mathscr{S}_{0,\,j} \quad \textrm{for}\; 0\leq \mathscr{N}_{0,\,m} < N,\;0\leq  \mathscr{S}_{0,\,j} < M,\nonumber \\ \quad \quad 0\leq f < \mathrm{S}_{\mathrm{E}}. 
\end{align}
Thus, the prediction filter is populated with a valid encoded symbol sequence as given in (\ref{equation:sstate_contents}).

Now, given the supertrellis state $\mathscr{S}_{\textrm{ST},\,j}$ and the input digit $\mathscr{N}_{0,\,m}$ in (\ref{equation:super_final}), the next supertrellis state $\mathscr{S}_{\textrm{ST},\,h}$ can be obtained as follows:
\begin{align}
\mathscr{F}_l :& \lbrace \mathscr{N}_{0,\,m} \mathscr{N}_{1,\,m} \hdots \mathscr{N}_{P-1,\,m} \rbrace \nonumber \\
h=& f \times N^P + l \quad 0\leq f < S_{\textrm{E}}, \nonumber \\
& 0\leq l \leq N^P-1, \; 0\leq \mathscr{S}_{\textrm{ST},\,h} \leq S_{\textrm{E}}\times N^P-1.
\end{align}
To summarize
\begin{align}
\mathscr{S}_{{\mathrm{ST}},\,j},\mathscr{N}_{0,\,m} \rightarrow \mathscr{S}_{{\mathrm{ST}},\,h},\mathscr{S}_{0,\,j}
\end{align}
which means: the supertrellis state $\mathscr{S}_{\mathrm{ST},\,j}$ with input digit $\mathscr{N}_{0,\,m}$ gives the code digit $\mathscr{S}_{0,\,j}$ and the next supertrellis state $\mathscr{S}_{\mathrm{ST},\,h}$. Note also that according to the notation in (\ref{super_sym_not})
\begin{align}
\mathscr{S}_{\textrm{ST},\,h}: \lbrace \mathscr{E}_f;\mathscr{F}_l\rbrace.
\end{align}

The supertrellis for a $1^{\textrm{st}}$-order predictor ($P=1$) and the rate-$1/2$ encoder given in Table \ref{tab:sim_par_lit_va} is given in Table \ref{tab:2}.
\subsection{The Predictive Viterbi algorithm \cite{vineel_va,branka,text_book}} 
\label{sec:pred_VA_ALG}
 Let $v_{k,\,m,\,n}$ denote the branch metric at time instant $k$ corresponding to the transition from state $m$ to state $n$. We have
\begin{align}
\label{branch_metric_va}
v_{k,\,m,\,n} &= \sum \limits_{l=0}^{N_r-1} \left| \tilde{z}_{k,\,m,\,n,\,l}\right|^2 \nonumber \\
                  &= \sum \limits_{l=0}^{N_r-1} \left| \sum \limits _{j=0}^P \tilde{a}_{P,\,j}\tilde{Y}_{k-j,\,l}/S_{j,\,m,\,n}\right|^2 
\end{align}
where $S_{0,\,m,\,n}$ denotes the input symbol corresponding to the transition from state $m$ to $n$ and the data $S_{j,\,m,\,n}$ are the contents of the prediction filter of state $m$. 

\subsubsection{Complexity Reduction using Isometry \cite{vin_wpc,elsevier,vineel_va}} \mbox{}
\label{comp_red_lit}
Consider the error signal $\tilde{z}_{k,\,m,\,n,\,l}$ in (\ref{branch_metric_va}): 
\begin{align}
\label{equation:complexity_reduction}
 \tilde{z}_{k,\,m,\,n,\,l} &= \tilde{X}_{k,\,l} - \hat{X}_{k,\,l} \nonumber \\ 
  &= \sum \limits^P_{j=0}\tilde{a}_{P,\,j} \frac{\tilde{Y}_{k-j,\,l}}{S_{j,\,m,\,n}}  \nonumber \\
  &=\frac{1}{S_{0,\,m,\,n}} \sum \limits^P_{j=0}\tilde{a}_{P,\,j} \frac{\tilde{Y}_{i-j,\,l}\times S_{0,\,m,\,n}}{S_{j,\,m,\,n}}.
 \end{align}
Note that $\left| \tilde{z}_{k,\,m,\,n,\,l} \right|^2$ is independent of $S_{0,\,m,\,n}$ (due to isometry \cite{elsevier}) and is dependent only on the phase changes between $S_{j,\,m,\,n}$ and $S_{0,\,m,\,n}$. In particular, the all-zero and all-one sequence $g_k$, yield the same magnitude squared error $\left| \tilde{z}_{k,\,m,\,n,\,l} \right|^2$, and are hence indistinguishable. 
 
 In general it is clear from (\ref{equation:complexity_reduction}) that two symbol sequences
\begin{align}
\mathbf{S}^{\left(\nu\right)}=\lbrace \hdots S^{\left(\nu\right)}_{k-1} S^{\left(\nu\right)}_{k} S^{\left(\nu\right)}_{k+1} \hdots \rbrace
\end{align} 
and
\begin{align}
\mathbf{S}^{\left(\omega\right)}=\lbrace \hdots S^{\left(\omega\right)}_{k-1} S^{\left(\omega\right)}_{k} S^{\left(\omega\right)}_{k+1} \hdots \rbrace
\end{align}  
are isometric if
\begin{align}
\mathbf{S}^{\left(\omega\right)} = e^{\mathrm{j}\phi}\mathbf{S}^{\left(\nu\right)}
\end{align}
where $\phi$ is a constant phase. This implies that we need to differentially encode $b_k$ at the transmitter. However, when differential encoding is done then $\mathscr{M}\left(\mathscr{S}_k \right) \neq S_k$ (see Figure \ref{fig:supertrellis} and Section \ref{section:supertrellis}). 
 
 Now consider (\ref{equation:complexity_reduction}). Note that $S_{0,\,m,\,n}/S_{1,\,m,\,n}$ is a function of the input code digit $\mathscr{S}_{0,\,j}$ in Figure \ref{fig:supertrellis} (see also Table \ref{tab:dif_lit1}). Mathematically, this can be stated as
\begin{align}
\frac{S_{0,\,m,\,n}}{S_{1,\,m,\,n}} = f_1\left(\mathscr{S}_{0,\,j} \right).
\end{align}
Similarly in (\ref{equation:complexity_reduction})
\begin{align}
\frac{S_{0,\,m,\,n}}{S_{2,\,m,\,n}} = f_2\left(\mathscr{S}_{0,\,j},\,\mathscr{S}_{1,\,j} \right)
\end{align} 
where $f_2\left(\mathscr{S}_{0,\,j},\mathscr{S}_{1,\,j} \right)$ is some function of $\mathscr{S}_{0,\,j}$ and $\mathscr{S}_{1,\,j}$ in Figure \ref{fig:supertrellis}, depending on the differential encoding rules in Table \ref{tab:dif_lit1}. Continuing in this manner we find that 
\begin{align}
\label{equation:complexity_justification}
\frac{S_{0,\,m,\,n}}{S_{P,\,m,\,n}} = f_P\left(\mathscr{S}_{0,\,j},\,\mathscr{S}_{1,\,j},\hdots,\mathscr{S}_{P-1,\,j} \right).
\end{align}
Thus, we find from (\ref{equation:complexity_justification}) that the metric in (\ref{equation:complexity_reduction}) is a function of only $P-1$ digits in the memory, with $\mathscr{S}_{0,\,j}$ being the input digit.  Thus the number of states in the trellis with differential encoding is only  $ M^{P-1}$ instead of $M^{P}$.  
 
The VA operates as follows. Let ${\cal C}_n$ denote the set of states that converge to the state $n$ ($0 \leq n \leq S_{\textrm{ST}}-1$). Let $\mu _{k,\,n}$ denote the path metric at time instant $k$ ($0 \leq k \leq L_d -1$) and state $n$. 
 \begin{enumerate}
\item {\tt Set initial values as $k=0$ and $\mu_{0,\,n}=0$ $\left(0\leq n \leq S_{\textrm{ST}}-1\right)$, since we assume that the receiver does not know the starting state}. 
\item  {\tt Increase time $k$ by 1.}
\item \label{main_step} {\tt Compute the path metrics at each state $n$ as }
 \begin{align}
\mu _{k,\,n} = \mathop {\mathrm{min}}\limits_{m \in {\cal C}_n}\;\left\lbrace v_{k,\,m,\,n} + \mu _{k-1,\,m} \right\rbrace.
\end{align}
\begin{enumerate}
\item {\tt Store the survivor for each state $n$}.
\item {\tt Identify the state having the minimum} $\mu _{k,\,n}${\tt , and trace back \newline along the survivor path and release a symbol corresponding \newline to time $k-{\cal D}^{'}_v$, where ${\cal D}^{'}_v$ is the decoding delay of the VA}.
\item {\tt Increase $k$ by 1.}
\item {\tt Go to step \ref{main_step} until time $k=L_d$}.
\end{enumerate}
\end{enumerate}  

\section{Simulation Results}
Since the overall rate is $1$, the average SNR per bit for each receive diversity arm is defined as
\cite{shimla,wpc}: 
\begin{align}
\textrm{SNR per bit} &= \frac{N_r \times {E}{\left[\left| \tilde H_{k,\,l} S_{k} \right|^2 \right]}}{{E}{\left[\left|\tilde W_{k,\,l}\right|^{2}\right]}} \nonumber \\
&= \frac{ N_r \times \left(L_h \times {2\sigma_{f}^{2}}\right) \times \left| S_k\right|^2}{L_d \times {2\sigma_{w}^{2}}}.
\end{align}
\begin{table}[H]
\fontsize{10}{10}\selectfont 
\centering
\caption{Simulation parameters}
\label{tab:sim_par_lit_va}       
\begin{tabular}{ll}
\toprule
Parameter & Value \\
\midrule
Frame size $L_d$ & 1024  \\
Channel memory $L_h-1$ & 9   \\
Length of the cyclic prefix $L_{\textrm{CP}}$ & 9 \\
Decoding delay of the VA ${\cal D}^{'}_v$ & 30 \\
No. of frames simulated & $ 10^{5}$  \\
Receiver antennas & 4  \\
1D channel fade variance $\sigma_f^2$ & 0.5 \\
Generator matrix for the encoder  &  $ \left[ 1 \quad \frac{1+D^2}{1+D+D^2} \right]$  \\
\bottomrule
\end{tabular}
\end{table}

\begin{table}[H]
\fontsize{10}{10}\selectfont 
\centering
\caption{Unnormalized supertrellis for the inner code, for a first-order ($P=1$) prediction order and a rate-$1/2$ encoder given in Table \ref{tab:sim_par_lit_va}}
\label{tab:2}       
\begin{tabular}{ccc}
\toprule
\makecell{Present supertrellis \\ state $j$ (time $n$)} & \makecell{Input $\mathscr{N}_{0,\,m}$} & \makecell{Next supertrellis \\ state $h$ (time $n+1$)}\\
\midrule
0    & 0  & 0  \\
0    & 1  & 5  \\
1    & 0  & 0  \\
1    & 1  & 5  \\         
2    & 0  & 4  \\
2    & 1  & 1  \\
3    & 0  & 4  \\          
3    & 1  & 1  \\
4    & 0  & 6  \\
4    & 1  & 3  \\          
5    & 0  & 6  \\
5    & 1  & 3  \\
6    & 0  & 2  \\
6    & 1  & 7  \\          
7    & 0  & 2  \\
7    & 1  & 7  \\
\bottomrule
\end{tabular}
\end{table}

\begin{figure}
\pgfplotsset{width=0.5\textwidth, height=0.5\textwidth}
\begin{center}
 \begin{tikzpicture}[scale=1]
    \begin{axis}[ymode=log,
        xlabel=SNR per bit (dB) ,
        ylabel=BER, legend cell align=left] 
    \addplot[mark=pentagon,mark options={scale=2,solid},black,thick] plot coordinates {
        (0,0.1269034)
        (4,0.0221382)
        (8,0.001557)
        (12,0.0000579)
        (14,0.0000111)

    };
    \addplot[black,mark=square,mark options={scale=2,solid}]
        plot coordinates {
        (0,0.359)
        (4,0.1865)
        (8,0.03572)
        (12,0.00246 )
        (17,0.0000418)

        };

      \addplot[dash pattern=on 1pt off 4pt on 4pt off 4pt, black,mark=triangle,mark options={scale=2,solid}]
        plot coordinates {
       (0,0.33518)
        (4,0.12656)
        (8,0.0162)
        (12,0.0009408 )
        (17,0.000027)
        };
        
         \addplot[ black,mark=oplus,mark options={scale=2,solid}]
        plot coordinates {
      (0,0.311)
        (4,0.0966)
        (8,0.0102121)
        (12,0.0005193 )
        (17,0.0000084)
        };

   \legend{ideal coherent \\$P=1,\,S_{\textrm{ST}}=4$\\$P=2,\,S_{\textrm{ST}}=8$\\$P=3,\,S_{\textrm{ST}}=16$\\}
    \end{axis}
    \end{tikzpicture}  
\caption{BER performance of linear prediction-based detection of convolutional coded OFDM system using the predictive Viterbi algorithm. }
\label{fig:lit_pred_va_res}
 \end{center} 
\end{figure}
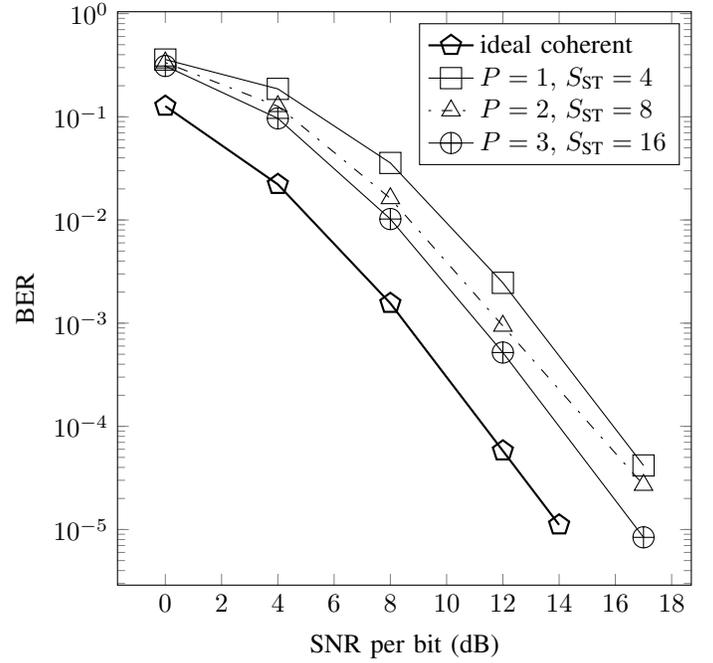
The simulation parameters are given in Table \ref{tab:sim_par_lit_va}. The BER performance of the linear prediction-based data detection in convolutional coded OFDM systems using predictive VA is given in Figure \ref{fig:lit_pred_va_res}. The LP-based receiver for convolutional coded SIMO-OFDM performs close to the ideal coherent receiver.
\section{Conclusion}
We have shown that the LP-based receiver performs close to the ideal coherent receiver. Future work can be focused on increasing the bit rate using $M$-ary constellations.

\ifCLASSOPTIONcaptionsoff
 \newpage
\fi
\bibliographystyle{IEEEtran}
	\bibliography{thesis}

%
%

%
%

\end{document}

%% file: predictive_va.pstex_t
\begin{picture}(0,0)%
\includegraphics{predictive_va.pstex}%
\end{picture}%
\setlength{\unitlength}{3947sp}%
\begingroup\makeatletter\ifx\SetFigFont\undefined%
\gdef\SetFigFont#1#2#3#4#5{%
  \reset@font\fontsize{#1}{#2pt}%
  \fontfamily{#3}\fontseries{#4}\fontshape{#5}%
  \selectfont}%
\fi\endgroup%
\begin{picture}(7965,2188)(-1073,-1775)
\put(3076,239){\makebox(0,0)[b]{\smash{{\SetFigFont{11}{13.2}{\rmdefault}{\mddefault}{\updefault}{\color[rgb]{0,0,0}$S_k$}%
}}}}
\put(1651,239){\makebox(0,0)[b]{\smash{{\SetFigFont{11}{13.2}{\rmdefault}{\mddefault}{\updefault}{\color[rgb]{0,0,0}$b_k$}%
}}}}
\put(-149,239){\makebox(0,0)[b]{\smash{{\SetFigFont{11}{13.2}{\rmdefault}{\mddefault}{\updefault}{\color[rgb]{0,0,0}$g_k$}%
}}}}
\put(-224,-811){\makebox(0,0)[b]{\smash{{\SetFigFont{11}{13.2}{\rmdefault}{\mddefault}{\updefault}{\color[rgb]{0,0,0}$\hat{g}_k$}%
}}}}
\put(4770,-997){\makebox(0,0)[b]{\smash{{\SetFigFont{10}{12.0}{\rmdefault}{\mddefault}{\updefault}{\color[rgb]{0,0,0}+}%
}}}}
\put(1951,-886){\makebox(0,0)[b]{\smash{{\SetFigFont{11}{13.2}{\rmdefault}{\mddefault}{\updefault}{\color[rgb]{0,0,0}$L_d$-pt.}%
}}}}
\put(-749, 89){\makebox(0,0)[b]{\smash{{\SetFigFont{11}{13.2}{\rmdefault}{\mddefault}{\updefault}{\color[rgb]{0,0,0}source}%
}}}}
\put(826,-136){\makebox(0,0)[b]{\smash{{\SetFigFont{12}{14.4}{\rmdefault}{\mddefault}{\updefault}{\color[rgb]{0,0,0}encoder}%
}}}}
\put(826,239){\makebox(0,0)[b]{\smash{{\SetFigFont{12}{14.4}{\rmdefault}{\mddefault}{\updefault}{\color[rgb]{0,0,0}rate-$r_1/r_2$}%
}}}}
\put(2401,-61){\makebox(0,0)[b]{\smash{{\SetFigFont{11}{13.2}{\rmdefault}{\mddefault}{\updefault}{\color[rgb]{0,0,0}mapper}%
}}}}
\put(3526,239){\makebox(0,0)[b]{\smash{{\SetFigFont{11}{13.2}{\rmdefault}{\mddefault}{\updefault}{\color[rgb]{0,0,0}S}%
}}}}
\put(3976,164){\makebox(0,0)[b]{\smash{{\SetFigFont{11}{13.2}{\rmdefault}{\mddefault}{\updefault}{\color[rgb]{0,0,0}$L_d$-pt.}%
}}}}
\put(5476,164){\makebox(0,0)[b]{\smash{{\SetFigFont{11}{13.2}{\rmdefault}{\mddefault}{\updefault}{\color[rgb]{0,0,0}insert}%
}}}}
\put(5476,-61){\makebox(0,0)[b]{\smash{{\SetFigFont{11}{13.2}{\rmdefault}{\mddefault}{\updefault}{\color[rgb]{0,0,0}CP}%
}}}}
\put(6001,-811){\makebox(0,0)[b]{\smash{{\SetFigFont{11}{13.2}{\rmdefault}{\mddefault}{\updefault}{\color[rgb]{0,0,0}Rayleigh}%
}}}}
\put(6001,-961){\makebox(0,0)[b]{\smash{{\SetFigFont{11}{13.2}{\rmdefault}{\mddefault}{\updefault}{\color[rgb]{0,0,0}frequency-selective}%
}}}}
\put(4876,-1711){\makebox(0,0)[b]{\smash{{\SetFigFont{11}{13.2}{\rmdefault}{\mddefault}{\updefault}{\color[rgb]{0,0,0}AWGN $\tilde{w}_{k,\,l}$}%
}}}}
\put(3676,-1111){\makebox(0,0)[b]{\smash{{\SetFigFont{11}{13.2}{\rmdefault}{\mddefault}{\updefault}{\color[rgb]{0,0,0}transient samples}%
}}}}
\put(3676,-961){\makebox(0,0)[b]{\smash{{\SetFigFont{11}{13.2}{\rmdefault}{\mddefault}{\updefault}{\color[rgb]{0,0,0}CP, }%
}}}}
\put(3676,-811){\makebox(0,0)[b]{\smash{{\SetFigFont{11}{13.2}{\rmdefault}{\mddefault}{\updefault}{\color[rgb]{0,0,0}remove}%
}}}}
\put(2401,-1036){\makebox(0,0)[b]{\smash{{\SetFigFont{11}{13.2}{\rmdefault}{\mddefault}{\updefault}{\color[rgb]{0,0,0}/}%
}}}}
\put(2401,-1186){\makebox(0,0)[b]{\smash{{\SetFigFont{11}{13.2}{\rmdefault}{\mddefault}{\updefault}{\color[rgb]{0,0,0}P}%
}}}}
\put(1951,-1111){\makebox(0,0)[b]{\smash{{\SetFigFont{11}{13.2}{\rmdefault}{\mddefault}{\updefault}{\color[rgb]{0,0,0}FFT}%
}}}}
\put(1126,-811){\makebox(0,0)[b]{\smash{{\SetFigFont{11}{13.2}{\rmdefault}{\mddefault}{\updefault}{\color[rgb]{0,0,0}$\tilde{Y}_k$}%
}}}}
\put(451,-886){\makebox(0,0)[b]{\smash{{\SetFigFont{11}{13.2}{\rmdefault}{\mddefault}{\updefault}{\color[rgb]{0,0,0}predictive }%
}}}}
\put(-749,-961){\makebox(0,0)[b]{\smash{{\SetFigFont{10}{12.0}{\rmdefault}{\mddefault}{\updefault}{\color[rgb]{0,0,0}sink}%
}}}}
\put(3526, 88){\makebox(0,0)[b]{\smash{{\SetFigFont{11}{13.2}{\rmdefault}{\mddefault}{\updefault}{\color[rgb]{0,0,0}/}%
}}}}
\put(1501,-811){\makebox(0,0)[b]{\smash{{\SetFigFont{11}{13.2}{\rmdefault}{\mddefault}{\updefault}{\color[rgb]{0,0,0}P}%
}}}}
\put(2401,-811){\makebox(0,0)[b]{\smash{{\SetFigFont{11}{13.2}{\rmdefault}{\mddefault}{\updefault}{\color[rgb]{0,0,0}S}%
}}}}
\put(4426,239){\makebox(0,0)[b]{\smash{{\SetFigFont{11}{13.2}{\rmdefault}{\mddefault}{\updefault}{\color[rgb]{0,0,0}P}%
}}}}
\put(4426,-136){\makebox(0,0)[b]{\smash{{\SetFigFont{11}{13.2}{\rmdefault}{\mddefault}{\updefault}{\color[rgb]{0,0,0}S}%
}}}}
\put(3526,-136){\makebox(0,0)[b]{\smash{{\SetFigFont{11}{13.2}{\rmdefault}{\mddefault}{\updefault}{\color[rgb]{0,0,0}P}%
}}}}
\put(2401,164){\makebox(0,0)[b]{\smash{{\SetFigFont{11}{13.2}{\rmdefault}{\mddefault}{\updefault}{\color[rgb]{0,0,0}DQPSK}%
}}}}
\put(3976,-61){\makebox(0,0)[b]{\smash{{\SetFigFont{11}{13.2}{\rmdefault}{\mddefault}{\updefault}{\color[rgb]{0,0,0}IFFT}%
}}}}
\put(451,-1111){\makebox(0,0)[b]{\smash{{\SetFigFont{11}{13.2}{\rmdefault}{\mddefault}{\updefault}{\color[rgb]{0,0,0}VA}%
}}}}
\put(6001,-1186){\makebox(0,0)[b]{\smash{{\SetFigFont{11}{13.2}{\rmdefault}{\mddefault}{\updefault}{\color[rgb]{0,0,0}fading channel $\tilde{h}_{k,\,l}$}%
}}}}
\put(895, 59){\makebox(0,0)[b]{\smash{{\SetFigFont{12}{14.4}{\rmdefault}{\mddefault}{\updefault}{\color[rgb]{0,0,0}convolutional}%
}}}}
\end{picture}%

%% file: gray_arxiv.pstex_t
\begin{picture}(0,0)%
\includegraphics{gray_arxiv.pstex}%
\end{picture}%
\setlength{\unitlength}{4144sp}%
\begingroup\makeatletter\ifx\SetFigFont\undefined%
\gdef\SetFigFont#1#2#3#4#5{%
  \reset@font\fontsize{#1}{#2pt}%
  \fontfamily{#3}\fontseries{#4}\fontshape{#5}%
  \selectfont}%
\fi\endgroup%
\begin{picture}(3239,2789)(1104,-4268)
\put(1891,-2536){\makebox(0,0)[b]{\smash{{\SetFigFont{11}{13.2}{\rmdefault}{\mddefault}{\updefault}{\color[rgb]{0,0,0}$\mathscr{M}\left(2\right)=-1+\mathrm{j}$}%
}}}}
\put(1666,-1996){\makebox(0,0)[b]{\smash{{\SetFigFont{11}{13.2}{\rmdefault}{\mddefault}{\updefault}{\color[rgb]{0,0,0}$10$}%
}}}}
\put(3781,-1996){\makebox(0,0)[b]{\smash{{\SetFigFont{11}{13.2}{\rmdefault}{\mddefault}{\updefault}{\color[rgb]{0,0,0}$00$}%
}}}}
\put(3781,-3931){\makebox(0,0)[b]{\smash{{\SetFigFont{11}{13.2}{\rmdefault}{\mddefault}{\updefault}{\color[rgb]{0,0,0}$01$}%
}}}}
\put(3601,-2536){\makebox(0,0)[b]{\smash{{\SetFigFont{11}{13.2}{\rmdefault}{\mddefault}{\updefault}{\color[rgb]{0,0,0}$\mathscr{M}\left(0\right)=1+\mathrm{j}$}%
}}}}
\put(3601,-3346){\makebox(0,0)[b]{\smash{{\SetFigFont{11}{13.2}{\rmdefault}{\mddefault}{\updefault}{\color[rgb]{0,0,0}$\mathscr{M}\left(1\right)=1-\mathrm{j}$}%
}}}}
\put(1891,-3346){\makebox(0,0)[b]{\smash{{\SetFigFont{11}{13.2}{\rmdefault}{\mddefault}{\updefault}{\color[rgb]{0,0,0}$\mathscr{M}\left(3\right)=-1-\mathrm{j}$}%
}}}}
\put(1666,-3931){\makebox(0,0)[b]{\smash{{\SetFigFont{11}{13.2}{\rmdefault}{\mddefault}{\updefault}{\color[rgb]{0,0,0}$11$}%
}}}}
\end{picture}%

%% file: Fig4_arxiv.pstex_t
\begin{picture}(0,0)%
\includegraphics{Fig4_arxiv.pstex}%
\end{picture}%
\setlength{\unitlength}{4144sp}%
\begingroup\makeatletter\ifx\SetFigFont\undefined%
\gdef\SetFigFont#1#2#3#4#5{%
  \reset@font\fontsize{#1}{#2pt}%
  \fontfamily{#3}\fontseries{#4}\fontshape{#5}%
  \selectfont}%
\fi\endgroup%
\begin{picture}(5984,1249)(2589,-3053)
\put(7291,-2716){\makebox(0,0)[b]{\smash{{\SetFigFont{12}{14.4}{\rmdefault}{\mddefault}{\updefault}{\color[rgb]{0,0,0}$\hdots$}%
}}}}
\put(7246,-1951){\makebox(0,0)[b]{\smash{{\SetFigFont{11}{13.2}{\rmdefault}{\mddefault}{\updefault}{\color[rgb]{0,0,0}prediction filter memory}%
}}}}
\put(4231,-2806){\makebox(0,0)[b]{\smash{{\SetFigFont{11}{13.2}{\rmdefault}{\mddefault}{\updefault}{\color[rgb]{0,0,0}convolutional encoder}%
}}}}
\put(3016,-2716){\makebox(0,0)[b]{\smash{{\SetFigFont{11}{13.2}{\rmdefault}{\mddefault}{\updefault}{\color[rgb]{0,0,0}$r_1$ bits $\vdots$}%
}}}}
\put(5581,-2896){\makebox(0,0)[b]{\smash{{\SetFigFont{11}{13.2}{\rmdefault}{\mddefault}{\updefault}{\color[rgb]{0,0,0}$=S_{0,\,j,\,h}$}%
}}}}
\put(8146,-2896){\makebox(0,0)[b]{\smash{{\SetFigFont{11}{13.2}{\rmdefault}{\mddefault}{\updefault}{\color[rgb]{0,0,0}$=S_{P,\,j,\,h}$}%
}}}}
\put(6526,-2896){\makebox(0,0)[b]{\smash{{\SetFigFont{11}{13.2}{\rmdefault}{\mddefault}{\updefault}{\color[rgb]{0,0,0}$=S_{1,\,j,\,h}$}%
}}}}
\put(5581,-2401){\makebox(0,0)[b]{\smash{{\SetFigFont{11}{13.2}{\rmdefault}{\mddefault}{\updefault}{\color[rgb]{0,0,0}$\mathscr{M}\left( \mathscr{S}_{0,\,j} \right)$}%
}}}}
\put(6531,-2671){\makebox(0,0)[b]{\smash{{\SetFigFont{11}{13.2}{\rmdefault}{\mddefault}{\updefault}{\color[rgb]{0,0,0}$\mathscr{M}\left( \mathscr{S}_{1,\,j}\right)$}%
}}}}
\put(8083,-2671){\makebox(0,0)[b]{\smash{{\SetFigFont{11}{13.2}{\rmdefault}{\mddefault}{\updefault}{\color[rgb]{0,0,0}$\mathscr{M}\left( \mathscr{S}_{P,\,j}\right)$}%
}}}}
\put(4231,-2536){\makebox(0,0)[b]{\smash{{\SetFigFont{11}{13.2}{\rmdefault}{\mddefault}{\updefault}{\color[rgb]{0,0,0}rate - $r_1/r_2$}%
}}}}
\end{picture}%

%% file: main.bbl
\begin{thebibliography}{10}
\providecommand{\url}[1]{#1}
\csname url@samestyle\endcsname
\providecommand{\newblock}{\relax}
\providecommand{\bibinfo}[2]{#2}
\providecommand{\BIBentrySTDinterwordspacing}{\spaceskip=0pt\relax}
\providecommand{\BIBentryALTinterwordstretchfactor}{4}
\providecommand{\BIBentryALTinterwordspacing}{\spaceskip=\fontdimen2\font plus
\BIBentryALTinterwordstretchfactor\fontdimen3\font minus
  \fontdimen4\font\relax}
\providecommand{\BIBforeignlanguage}[2]{{%
\expandafter\ifx\csname l@#1\endcsname\relax
\typeout{** WARNING: IEEEtran.bst: No hyphenation pattern has been}%
\typeout{** loaded for the language `#1'. Using the pattern for}%
\typeout{** the default language instead.}%
\else
\language=\csname l@#1\endcsname
\fi
#2}}
\providecommand{\BIBdecl}{\relax}
\BIBdecl

\bibitem{vin_wpc}
\BIBentryALTinterwordspacing
V.~K. Veludandi and K.~Vasudevan, ``Linear prediction-based detection of
  serially concatenated {DQPSK} in {SIMO}-{OFDM},'' \emph{Wireless Personal
  Communications}, Aug 2017. [Online]. Available:
  \url{https://doi.org/10.1007/s11277-017-4751-9}
\BIBentrySTDinterwordspacing

\bibitem{app_based2}
P.~Hoeher and J.~Lodge, ````{Turbo DPSK}' : {Iterative} differential {PSK}
  demodulation and channel decoding,'' \emph{IEEE Transactions on
  Communications}, vol.~47, no.~6, pp. 837--843, Jun 1999.

\bibitem{text_book}
K.~Vasudevan, \emph{{Digital {Communications} and {Signal} {Processing}, Second
  edition (CD-ROM included)}}.\hskip 1em plus 0.5em minus 0.4em\relax
  Universities Press (India), Hyderabad, 2010.

\bibitem{elsevier}
K.~Vasudevan, K.~Giridhar, and B.~Ramamurthi, ``Efficient suboptimum detectors
  based on linear prediction in {Rayleigh} flat-fading channels,'' \emph{Signal
  Processing Journal, Elsevier Science}, vol.~81, no.~4, pp. 819--828, April
  2001.

\bibitem{vineel_va}
\BIBentryALTinterwordspacing
V.~K. Veludandi and K.~Vasudevan, ``Noncoherent detection of {DQPSK} in {OFDM}
  systems using predictive {VA},'' in \emph{Journal of Physics: Conference
  Series}, vol. 787, no.~1.\hskip 1em plus 0.5em minus 0.4em\relax IOP
  Publishing, 2017, pp. 1--5. [Online]. Available:
  \url{http://stacks.iop.org/1742-6596/787/i=1/a=012024}
\BIBentrySTDinterwordspacing

\bibitem{pilot_based1}
B.~Yang, Z.~Cao, and K.~B. Letaief, ``Analysis of low-complexity windowed
  {DFT}-based {MMSE} channel estimator for {OFDM} systems,'' \emph{IEEE
  Transactions on Communications}, vol.~49, no.~11, pp. 1977--1987, Nov 2001.

\bibitem{pilot_based2}
S.~Coleri, M.~Ergen, A.~Puri, and A.~Bahai, ``Channel estimation techniques
  based on pilot arrangement in {OFDM} systems,'' \emph{IEEE Transactions on
  Broadcasting}, vol.~48, no.~3, pp. 223--229, Sep 2002.

\bibitem{pilot_based3}
M.~H. Hsieh and C.~H. Wei, ``Channel estimation for {OFDM} systems based on
  comb-type pilot arrangement in frequency selective fading channels,,''
  \emph{IEEE Transactions on Consumer Electronics}, vol.~44, no.~1, pp.
  217--225, Feb 1998.

\bibitem{super_trellis1}
M.~V. Eyuboglu and S.~U.~H. Qureshi, ``Reduced-state sequence estimation for
  coded modulation of intersymbol interference channels,'' \emph{IEEE Journal
  on Selected Areas in Communications}, vol.~7, no.~6, pp. 989--995, Aug 1989.

\bibitem{super_trellis2}
P.~R. Chevillat and E.~Eleftheriou, ``Decoding of trellis-encoded signals in
  the presence of intersymbol interference and noise,'' \emph{IEEE Transactions
  on Communications}, vol.~37, no.~7, pp. 669--676, Jul 1989.

\bibitem{sivp}
K.~Vasudevan, ``Turbo equalization of serially concatenated turbo codes using a
  {Predictive}-{DFE} based receiver,'' \emph{Signal, Image and Video
  Processing, Springer}, vol.~1, no.~3, pp. 239--252, August 2007.

\bibitem{text_book_soft}
\BIBentryALTinterwordspacing
------, ``Digital communications and signal processing, third edition,'' 2016,
  [Online; accessed 8-February-2016]. [Online]. Available:
  \url{https://home.iitk.ac.in/\~ vasu/book0.pdf}
\BIBentrySTDinterwordspacing

\bibitem{indouk}
\BIBentryALTinterwordspacing
------, ``Coherent detection of turbo-coded {OFDM} signals transmitted through
  frequency selective {Rayleigh} fading channels with receiver diversity and
  increased throughput,'' \emph{CoRR}, vol. abs/1511.00776, 2015. [Online].
  Available: \url{http://arxiv.org/abs/1511.00776}
\BIBentrySTDinterwordspacing

\bibitem{shimla}
------, ``Coherent detection of turbo coded {OFDM} signals transmitted through
  frequency selective {Rayleigh} fading channels,'' in \emph{2013 IEEE
  International Conference on Signal Processing, Computing and Control
  (ISPCC)}, Shimla, India, Sept 2013, pp. 1--6.

\bibitem{wpc}
------, ``Coherent detection of turbo-coded {OFDM} signals transmitted through
  frequency selective {Rayleigh} fading channels with receiver diversity and
  increased throughput,'' \emph{Wireless Personal Communications, Springer},
  vol.~82, no.~3, pp. 1623--1642, June 2015.

\bibitem{spain}
------, ``Coherent turbo coded {MIMO} {OFDM},'' in \emph{ICWMC 2016, The
  Twelfth International Conference on Wireless and Mobile Communications},
  Barcelona, Spain, Nov 2016, pp. 91--99.

\bibitem{branka}
B.~Vucetic and J.~Yuan, \emph{Turbo codes: {Principles} and
  applications}.\hskip 1em plus 0.5em minus 0.4em\relax Kluwer Academic
  Publishers, Norwell, MA, 2000.

\end{thebibliography}
